\documentclass{aip-cp}
\usepackage[numbers]{natbib}
\usepackage{rotating}
\usepackage{graphicx}
\usepackage{aas_macros} 
\newcommand{\rmd}{{\rm d}}

\begin{document}

\title{Degeneracy in Studying the Supranuclear Equation of State and Modified Gravity with Neutron Stars}

\author[aff1]{Lijing Shao\corref{cor1}\noteref{note1}}

\affil[aff1]{Kavli Institute for Astronomy and Astrophysics, Peking University,
Beijing 100871, China}
\corresp[cor1]{Corresponding author: lshao@pku.edu.cn}
\authornote[note1]{To appear in the AIP Conference Proceedings of the Xiamen-CUSTIPEN Workshop on the EOS of
Dense Neutron-Rich Matter in the Era of Gravitational Wave Astronomy (January 3
to 7, 2019; Xiamen China).}

\maketitle

\begin{abstract}
  It is generally acknowledged that an extrapolation in physics from a well-known scale to an unknown scale is perilous.
  This prevents us from using laboratory experience to gain {\it precise} information for the supranuclear matter inside neutron stars (NSs).
  With operating and upcoming astronomical facilities, NSs' equation of state (EOS) is expected to be determined at a new level in the near future, {\it under the assumption that general relativity (GR) is the correct theory for gravitation}. 
  While GR is a reasonable working assumption yet still an extrapolation, there could be a large uncertainty due to the not-so-well-tested strong gravitational field inside NSs. 
  Here we review some recent theoretical efforts towards a better understanding of the degeneracy between the supranuclear EOS and alternative gravity theories.
\end{abstract}

\section{INTRODUCTION}

Neutron stars (NSs) provide the best celestial laboratory to study the coupling between the strong-field gravity and the matter fields \cite{Wex:2014nva,Will:2014kxa, Berti:2015itd, Kramer:2016kwa, Abbott:2018lct}. 
In Einstein's general relativity (GR), it is postulated that such a coupling is minimal, and the dynamics of the spacetime and matter fields are derived from a simple and esthetically appealing action \cite{Misner:1974qy, Poisson:2014misc},
\begin{equation}
  \label{eq:action}
  S = \frac{1}{16\pi G} \int \rmd x^4 \sqrt{-g} \, R + S_{\rm matter} \left[ \psi; g_{\mu\nu} \right] \,,
\end{equation}
where $g$ is the determinant of the metric $g_{\mu\nu}$, $R$ is the Ricci scalar, and $S_{\rm matter}$ is the action for all the matter degrees which are collectively denoted as $\psi$. 
The strikingly neat insight in Eq.~(\ref{eq:action}) is the universal coupling in $S_{\rm matter}\left[ \psi; g_{\mu\nu} \right]$ between the matter degrees and the spacetime metric. 
It follows from the principle of equivalence \cite{Will:1993ns, Will:2014kxa, Shao:2016ezh}. Equipped with this principle, it is validate to couple matter fields to the spacetime via the principle of general covariance \cite{Misner:1974qy} thus obtaining the quantum fields on a classically curved spacetime.

The field equation derived from the action is \cite{Misner:1974qy},
\begin{equation}
  \label{eq:EE}
  R_{\mu\nu} - \frac{1}{2}g_{\mu\nu} R = 8\pi G T_{\mu\nu}^{\rm matter} \,,
\end{equation}
where $R_{\mu\nu}$ is the Ricci tensor and $T_{\mu\nu}^{\rm matter}$ is the energy-momentum tensor for matters.
The consequence of this equation is that, as nicely summarized by John A. Wheeler, ``matter tells spacetime how to curve, and spacetime tells matter how to move.'' 
In our everyday experience, the spacetime is hardly curved however, due to the smallness of the gravitational coupling constant ``$G$''. 
Only with exotic objects like the NSs, the spacetime reacts to the dense matter (namely, the $T_{\mu\nu}$) in a highly significant way.

Despite its delicate beauty, there are reasons to question GR, including the observational facts like the ``dark matter'' and the ``dark energy'', as well as the theoretical dilemmas like the unavoidable singularities and the black-hole information loss problem \cite{Will:2014kxa, AmelinoCamelia:2008qg}.
Naive attempts to modify the Einstein's equation (\ref{eq:EE}) are classified, unsurprisingly, into two categories: (i) modifying the left-hand side (i.e., the geometric property of spacetime) and (ii) modifying the right-hand side (i.e., the contents of the matter world).
These two approaches are, {\it somehow}, degenerate (see {\it e.g.}, Ref.~\cite{He:2014yqa}).
For example, in a phenomenological approach where an extra Yukawa term is augmented to the Newtonian potential mimicking a fifth force mediated by a new massive particle, the effect of such a modification is captured by modifying the equation of state (EOS) \cite{Wen:2009av, Wen:2011rb}. 
In such a case, the observed 2-solar-mass NSs can even be explained by the very soft EOSs that are unable to supported 2-solar-mass NSs in GR \cite{Wen:2009av, Wen:2011rb}.

Fortunately, degeneracy is not the whole story. 
If non-minimal couplings between spacetime and matters are allowed, there exist many more ways to modify the Einstein's equation \cite{Brans:1961sx, Will:1993ns, Will:2014kxa}. 
Nevertheless, with non-minimal couplings, the degeneracy between modified gravity and matter contents is not gone altogether.
We emphasize that,
especially when the supranuclear EOS for NSs [$T_{\mu\nu}$ in Eq.~(\ref{eq:EE})] is quite uncertain, and the alternative gravity theories in the strong field are not empirically examined thoroughly, the degeneracy should not be overlooked.
Though a complete picture in studying such a degeneracy is still lacking, here we review some efforts along this line.

The paper is organized as follows.
In the next section, we give the theoretical ingredients to obtain the NSs'
structure in GR. Then, keeping alternative gravity theories in mind, we review
the work to extend the framework with perturbative approaches and non-perturbative approaches. Some discussions are presented at the end.
Throughout the paper we use the unit system where the light speed $c=1$.

\section{THE STRUCTURE OF A NS IN GR}
\label{sec:GR}

We consider a NS that is made of perfect fluid with an energy-momentum tensor $T^{\mu\nu} = \left( \epsilon + p \right) u^\mu u^\nu + p g^{\mu\nu}$, where $u^\mu$ is the fluid element's four-velocity, $p$ and $\epsilon$ are pressure and energy density respectively. Under the assumption of spherical symmetry,
Tolman-Oppenheimer-Volkoff (TOV) equations describe a fully relativistic NS in hydrostatic equilibrium \cite{Tolman:1939jz, Oppenheimer:1939ne, Misner:1974qy},
\begin{eqnarray}
  \frac{\rmd p}{\rmd r} &=& - G \frac{\epsilon + p}{r^2} \frac{m + 4\pi r^3 p}{1 - 2Gm/r} \,, \label{eq:TOV1:GR} \\
  \frac{\rmd m}{\rmd r} &=& 4\pi r^2 \epsilon \,, \label{eq:TOV2:GR}
\end{eqnarray}
where $m$, $\epsilon$, and $p$ are functions of the stellar radius $r$.

\begin{figure}[h]
  \centerline{\includegraphics[width=15cm]{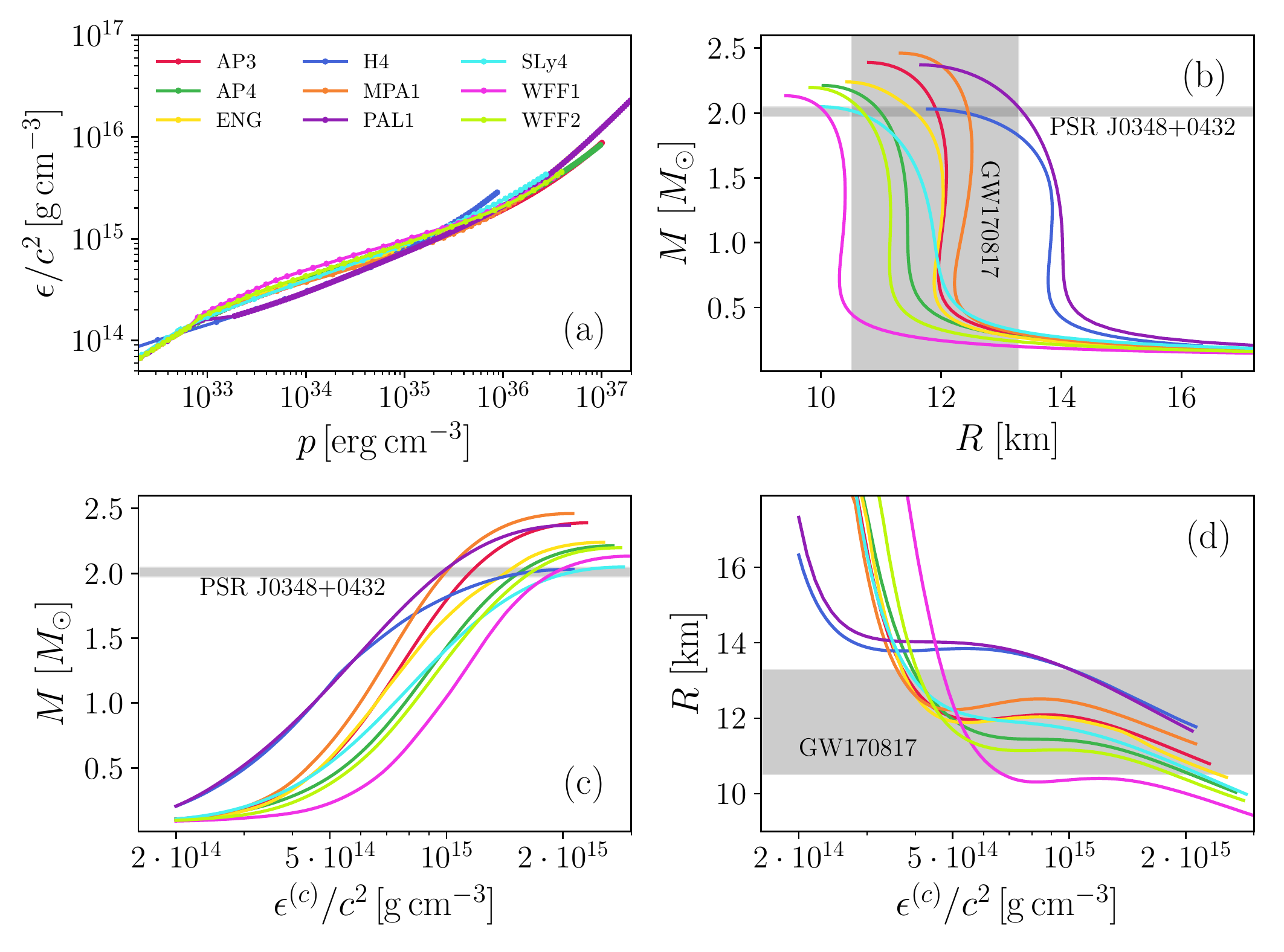}}
  
  \caption{\label{fig:NS:GR} NSs in GR with 9 EOSs. 
  {\bf (a)} 9 EOSs \cite{Lattimer:2000nx} that have a maximum mass for NSs  larger than 2\,$M_\odot$.
  {\bf (b)} Mass-radius relation for NSs.
  {\bf (c)} The mass of NSs as a function of the central energy density $\epsilon^{(c)}$.
  {\bf (d)} The radius of NSs as a function of the central energy density $\epsilon^{(c)}$. In (b) and (c) we have included the 1-$\sigma$ measurement for the mass of PSR~J0348+0432 \cite{Antoniadis:2013pzd}. In (b) and (d) we have included the measurement of the NS radii at 90\% confidence level from the binary NS inspiral GW170817 event \cite{TheLIGOScientific:2017qsa, Abbott:2018exr}. }
\end{figure}

Given an EOS, $\epsilon = \epsilon(p)$, and a central energy density $\epsilon^{(c)}$, the above equations can be integrated up to the stellar surface where $r=R$ and $p(R)=0$. For a variety of $\epsilon^{(c)}$ one obtains the mass-radius relation $M(R)$ where $M \equiv m(R)$ is the {\it Schwarzschild} mass (in alternative gravity theories it can be different from the Arnowitt-Deser-Misner mass). In Figure~\ref{fig:NS:GR} we present the calculation in GR for 9 EOSs \cite{Lattimer:2000nx} that have a maximum NS mass larger than 2\,$M_\odot$. As we can see, the current observations, from PSR~J0348+0432 \cite{Antoniadis:2013pzd} and GW170817 \cite{TheLIGOScientific:2017qsa, Abbott:2018exr}, are not capable to definitely pin down the correct EOS for NSs yet. However, if the measurements from GW170817 are taken into account, though with significant uncertainties, EOSs {\sf H4}, {\sf PAL1}, and {\sf WFF1} are starting to be in tension with observations \cite{Abbott:2018exr, Abbott:2018wiz}.

\section{MODIFIED GRAVITY: PERTURBATIVE REGIME}

In alternative gravity theories, the TOV equations for NSs are modified. There are in general two approaches to study the NS structures with modified gravity: {\it theory specific} or {\it generally parameterized}. 
\begin{itemize}
  \item In the former case, one usually needs to work out field
  configurations, as well as the metric and its dependence on the matter
  fields. In most situations, the metric is not the unique {\it gravitational
  field}, so one also has to worry about the spacetime profile of extra
  fields. Nevertheless, for a well-defined problem in a well-proposed gravity theory, one will obtain the fully
  relativistic nonlinear equations, similar to
  Eqs.~(\ref{eq:TOV1:GR}--\ref{eq:TOV2:GR}) in GR; see the next section for
  an example.
  \item In the latter case, the goal is to use plausible parameterization to
  cover as many alternative gravity theories as possible. Due to the
  nonlinearity inherent for the gravitational interaction, it becomes very
  hard, if ever possible, to incorporate all sensible effects with a limited
  number of free parameters. Nevertheless, it has the advantage of being {\it
  once for all}, and being advantageous from data analysis point of view.
\end{itemize}

In this section, we will review some earlier work to parameterize the TOV equation, starting from the famous parameterized post-Newtonian (PPN) framework \cite{Will:1993ns, Wagoner:1974ads,Ciufolini:1983ads} and a recently improved version, the post-TOV formalism \cite{Glampedakis:2015sua, Glampedakis:2016pes}. Both of them belong to the catalog of perturbative approach.

\subsection{Modified TOV equations in the PPN formalism}

The PPN formalism was originally developed to test alternative gravity theories in the Solar System in a systematic way \cite{Will:1972zz, Nordtvedt:1972zz, Will:1993ns}. Due to the intrinsic weak-field and slow-motion characteristics for bodies in the Solar System, the formalism was naturally adopting a post-Newtonian (PN) approximation, and only terms at 1\,PN order are included. Therefore, the PPN formalism captures (almost) all relativistic corrections to GR {\it at the leading PN order} in the Solar System. Nowadays, the free parameters in the PPN formalism are already well constrained by various observations \cite{Will:2014kxa, Wex:2014nva, Will:1993ns, Shao:2016ezh}.

\begin{figure}[h]
  \centerline{\includegraphics[width=9cm]{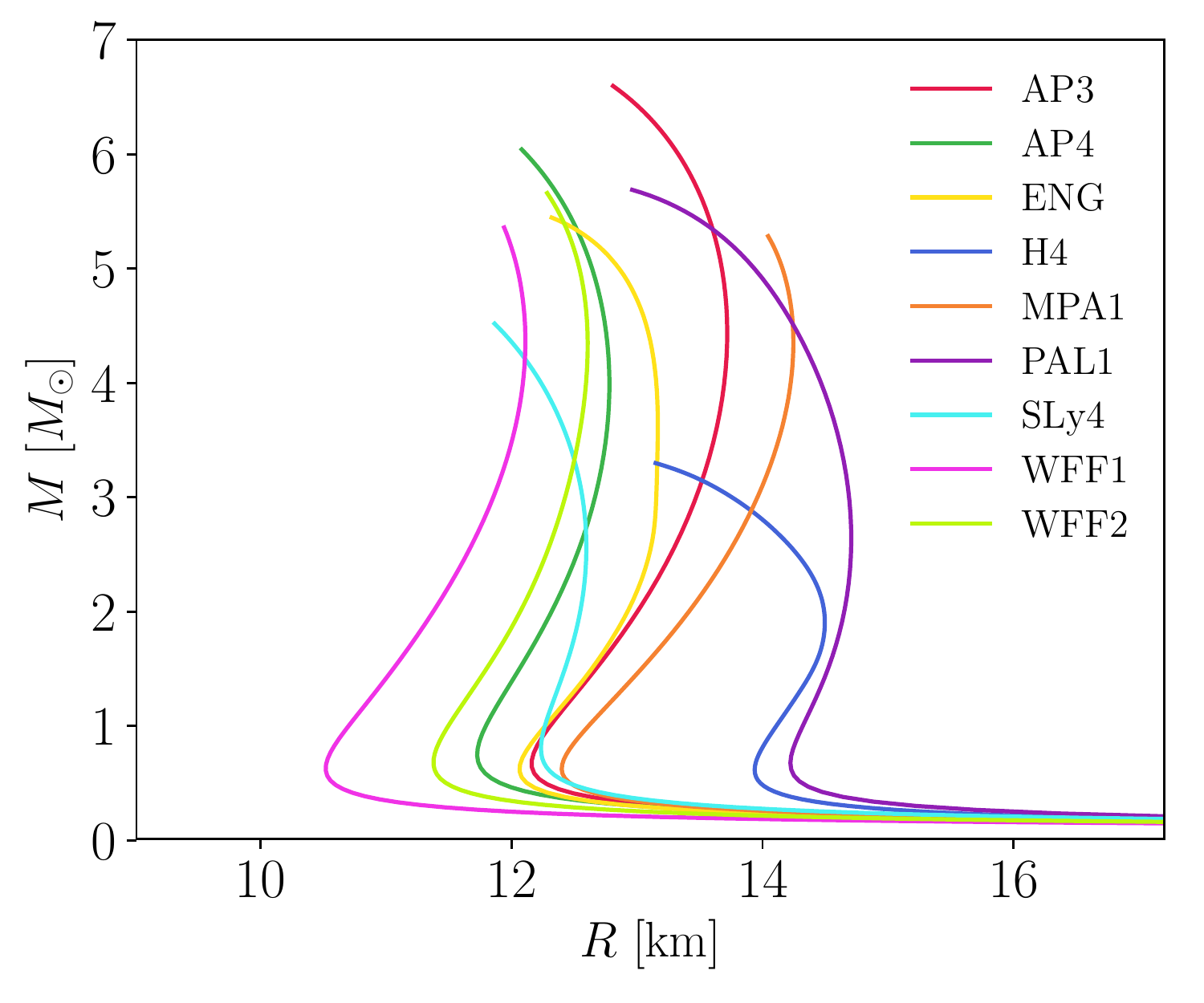}}
  
  \caption{\label{fig:NS:GR:1PN} ``Unphysical'' mass-radius relation for NSs from integrating the TOV equations in GR, but expanded to the 1\,PN order.}
\end{figure}

Based on the earlier work by \citet{Wagoner:1974ads} and \citet{Ciufolini:1983ads},  \citet{Glampedakis:2015sua} used an ``improved'' gauge choice which enables an easier comparison with the TOV equations in GR. They obtained a set of modified TOV equations at 1\,PN order,
\begin{eqnarray}
  \frac{\rmd p}{\rmd r} &=& - \frac{ Gm \rho}{r^2} \left[ 1 + \Pi + \frac{p}{\rho} + \left( 5+3\gamma-6\beta+\zeta_2 \right) \frac{Gm}{r} + \left( \gamma + \zeta_4 \right)  \frac{4\pi r^3 p}{m} \right] \,, \label{eq:TOV1:PPN} \\
  \frac{\rmd m}{\rmd r} &=& 4\pi r^2 \rho \left[ 1 + \left( 1+\zeta_3 \right) \Pi - \frac{1}{2} \left( 11+\gamma -12\beta + \zeta_2 -2\zeta_4 \right) \frac{Gm}{r} \right] \,. \label{eq:TOV2:PPN}
\end{eqnarray}
In the above equations, $\beta$, $\gamma$, and $\zeta_i$ are PPN parameters (in GR, $\beta=\gamma=1$ and $\zeta_i=0$; see Refs.~\cite{Will:1993ns, Will:2014kxa} for details), $\rho$ is the baryonic rest-mass density, and $\Pi \equiv \left( \epsilon - \rho \right) / \rho$. Eq.~(\ref{eq:TOV1:PPN}) and Eq.~(\ref{eq:TOV2:PPN}), when using the GR values for the PPN parameters, recover Eq.~(\ref{eq:TOV1:GR}) and Eq.~(\ref{eq:TOV2:GR}) respectively, if the latter set of equations were expanded to the 1\,PN order \cite{Glampedakis:2015sua}. 
Therefore, Eqs.~(\ref{eq:TOV1:PPN}--\ref{eq:TOV2:PPN}) are a generalization of the TOV equations {\it at 1\,PN} order. They shall be general enough for bodies that equipped with weak field and slow motion.

However, NSs have strong gravitational fields inside.
In Figure~\ref{fig:NS:GR:1PN}, we plot the mass-radius relation for NSs by integrating the TOV equations in GR, but expanded to 1\,PN order, or equivalently, by integrating Eqs.~(\ref{eq:TOV1:PPN}--\ref{eq:TOV2:PPN}) with PPN parameters set to their GR values.
It is easily seen that the results are ``unphysical'' {\it by a lot}, due to the omission of higher-order PN terms. 
{\it NSs are intrinsically strong-field objects!} 
Therefore, searching for modified-gravity signals based on the 1\,PN-expended TOV equations is not going to be useful.

\subsection{The post-TOV formalism}

Inspired by the PPN formalism \cite{Will:1972zz, Will:1993ns}, and to go beyond the leading-order PN approximation, a PN-nonlinear {\it hybrid} framework to study the structure of NSs was developed by \citet{Glampedakis:2015sua}. 
It is dubbed as the post-TOV formalism \cite{Glampedakis:2015sua,
Glampedakis:2016pes}. The framework collects the 1\,PN terms in
Eqs.~(\ref{eq:TOV1:PPN}--\ref{eq:TOV2:PPN}) that also appear in the 1\,PN
expansion of Eqs.~(\ref{eq:TOV1:GR}--\ref{eq:TOV2:GR}); these terms are {\it resumed}
to the GR form to mimic nonlinear effects as much as possible. The remaining
1\,PN terms are left as is (namely, 1\,PN Taylor expanded), forming the 1\,PN corrections to the TOV equations \cite{Glampedakis:2015sua}.

\begin{figure}[h]
  \centerline{\includegraphics[width=9cm]{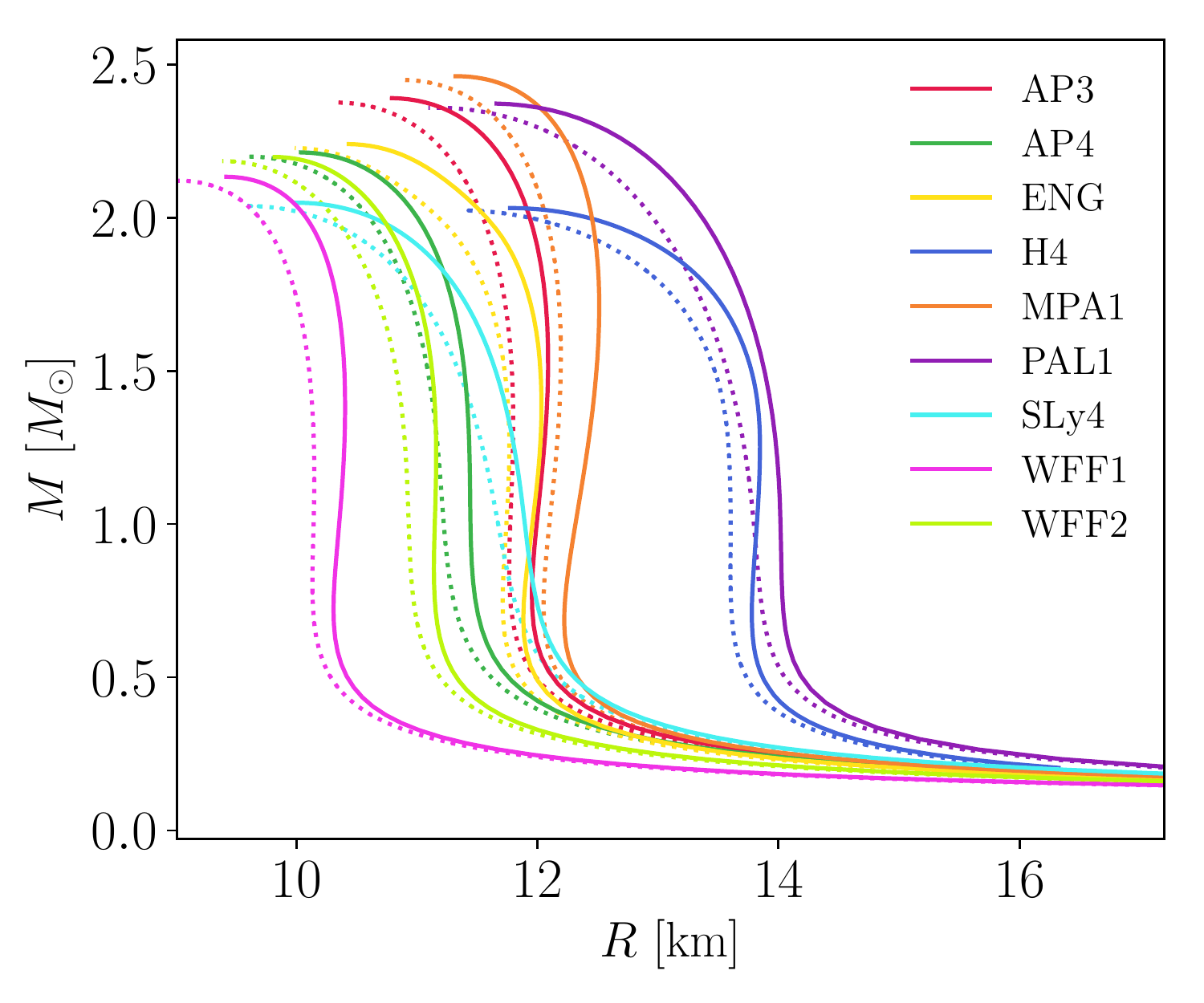}}
  \caption{\label{fig:NS:GR:postTOV} Mass-radius relation for NSs from
  integrating the TOV equations in GR (solid lines; same as panel (b) in
Figure~\ref{fig:NS:GR}) and post-TOV equations with (1\,PN parameter)
$\delta_3=0.5$ (dotted lines; notice that such a value for $\delta_3$ was
already excluded by Solar System and radio pulsar observations \cite{Will:2014kxa}); the other post-TOV parameters are set to zero.}
\end{figure}

\begin{figure}[h]
  \centerline{\includegraphics[width=9cm]{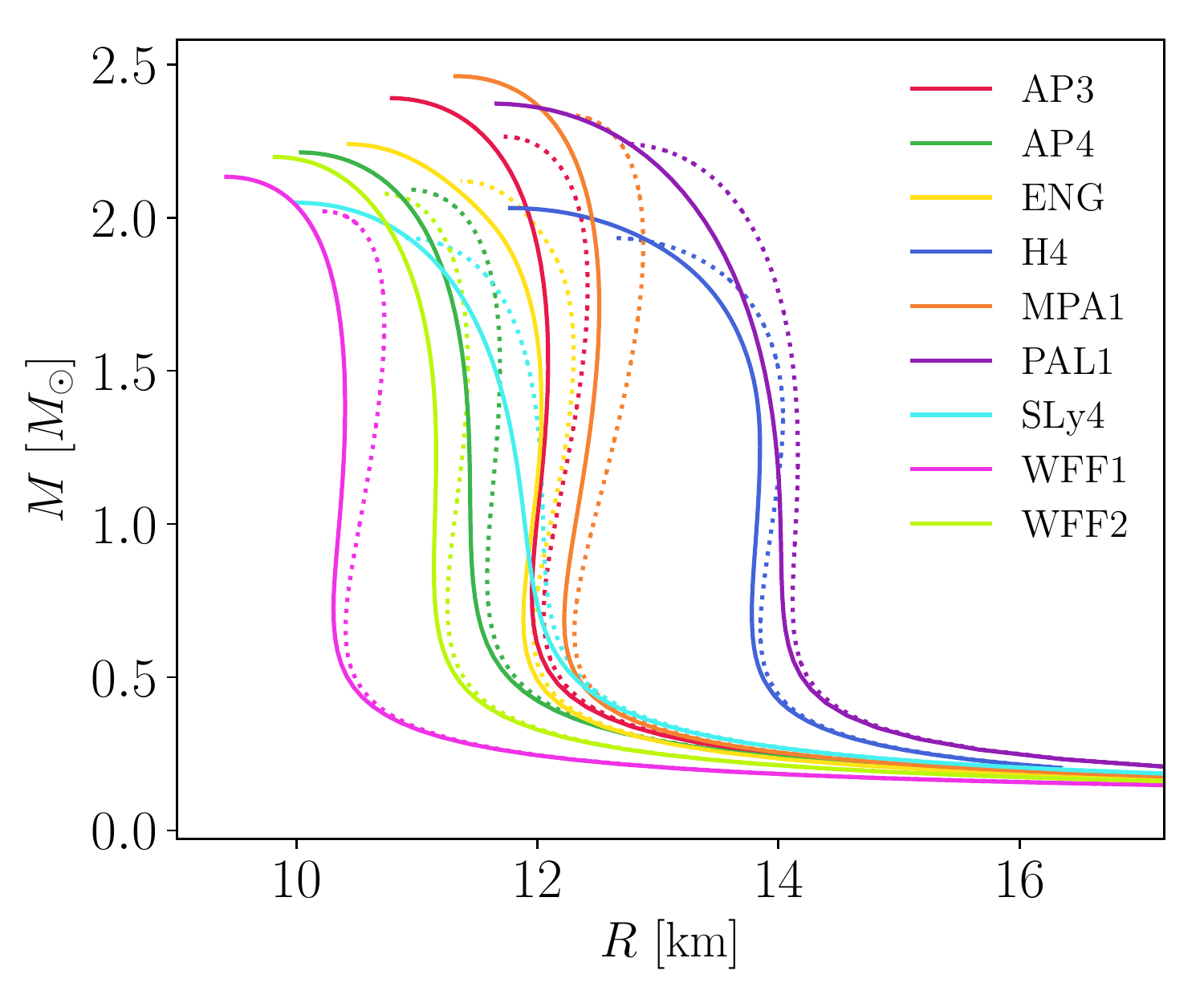}}
  \caption{\label{fig:NS:GR:postTOV2} Same as Figure~\ref{fig:NS:GR:postTOV}, but with (2\,PN parameter) $\mu_1=-0.5$ for the post-TOV case; the other post-TOV parameters are set to zero.}
\end{figure}

In addition to the 1\,PN corrections, with some reasonable assumptions,
\citet{Glampedakis:2015sua} also worked out the generic 2\,PN terms for the
post-TOV equations. Some simplifications were made, such as grouping self-similar 2\,PN functionals and so on (see the original paper for details). 
The final post-TOV equations at 2\,PN read \cite{Glampedakis:2015sua, Glampedakis:2016pes},
\begin{eqnarray}
  \frac{\rmd p}{\rmd r} &=& - G \frac{\epsilon + p}{r^2} \frac{m + 4\pi r^3 p}{1 - 2Gm/r} - \frac{G m \rho}{r^2} \left( {\cal P}_1 + {\cal P}_2 \right) \,, \label{eq:TOV1:postTOV} \\
  \frac{\rmd m}{\rmd r} &=& 4\pi r^2 \epsilon + 4\pi r^2 \rho \left( {\cal M}_1 + {\cal M}_2 \right) \,, \label{eq:TOV2:postTOV}
\end{eqnarray}
where the corrections are all encoded in ${\cal P}_i$ and ${\cal M}_i$ $(i=1,2)$. These corrections are \cite{Glampedakis:2015sua},
\begin{eqnarray}
  {\cal P}_1 &=& \delta_1 \frac{Gm}{r} +  \delta_2 \frac{4\pi r^3 p}{m} \,, \\
  {\cal M}_1 &=& \delta_3 \frac{Gm}{r} + \delta_4 \Pi  \,, \\
  {\cal P}_2 &=& \pi_1 \frac{G^2 m^3}{r^5 \rho} + \pi_2 \frac{G^2 m^2}{r^2} + \pi_3  G r^2 p + \pi_4 \frac{\Pi p}{\rho} \,, \\
  {\cal M}_2 &=& \mu_1 \frac{G^2 m^3}{r^5 \rho} + \mu_2 \frac{G^2 m^2}{r^2} + \mu_3 G r^2 p + \mu_4 \frac{\Pi p}{\rho} + \mu_5 \Pi^3 \frac{r}{G m} \,,
\end{eqnarray}
where $\delta_i$ $(i=1,\cdots,4)$, $\pi_i$ $(i=1,\cdots,4)$, and $\mu_i$ $(1,\cdots,5)$ are post-TOV parameters which vanish in GR. 

The 1\,PN correction terms, ${\cal P}_1$ and ${\cal M}_1$, are characterized by $\delta_i$'s which are simply linear combinations of the PPN parameters in Eqs.~(\ref{eq:TOV1:PPN}--\ref{eq:TOV2:PPN}) \cite{Glampedakis:2015sua},
\begin{eqnarray}
  \delta_1 &\equiv& 3\left( 1+\gamma \right) - 6\beta + \zeta_2 \,, \\
  \delta_2 &\equiv& \gamma - 1 + \zeta_4 \,, \\
  \delta_3 &\equiv& - \frac{1}{2} \left( 11 + \gamma - 12\beta + \zeta_2 -2\zeta_4 \right) \,, \\
  \delta_4 &\equiv& \zeta_3 \,.
\end{eqnarray}
Due to the tight constraints from Solar System and radio pulsars \cite{Wex:2014nva, Will:2014kxa, Shao:2012eg, Shao:2013wga, Shao:2013eka, Shao:2016ezh}, one has ${\cal P}_1 \ll 1$ and ${\cal M}_1 \ll 1$. In the limit that ${\cal P}_1 = {\cal M}_1 = 0$, the 2\,PN corrections can be {\it effectively} described by a gravity-modified energy density \cite{Glampedakis:2015sua, Glampedakis:2016pes} at leading order,
\begin{equation}
  \label{eq:epsilon:eff}
  \epsilon_{\rm eff} = \epsilon + \rho {\cal M}_2 \,.
\end{equation}
Therefore, in the post-TOV formalism the EOS and the gravity theory become degenerate. This conclusion agrees with a similar one in \citet{Wen:2009av} when a Yukawa correction to the Newtonian potential is considered, though, the post-TOV formalism assumes no massive propagating gravitational modes,
and it does not include exponentially suppressed correction like that of a Yukawa term at the first place (see Will's monograph \cite{Will:1993ns} for details).

In Figures~\ref{fig:NS:GR:postTOV} and \ref{fig:NS:GR:postTOV2}, we plot illustrative cases where we have fixed (1\,PN parameter) $\delta_3 = 0.5$ and (2\,PN parameter) $\mu_1 = -0.5$ respectively.\footnote{The value $\delta_3=0.5$ was already excluded from the observations from Solar System and radio pulsars \cite{Will:2014kxa}. Here we only use it as an illustrative example.} 
We see that, (i) compared with the naive PN expansion in the last subsection, now the behaviors of $M$-$R$ curves are much more {\it regulated} due to the inclusion of nonlinear effects {\it from GR} by the resummation; (ii) with varying post-TOV parameters, one is able to move the $M$-$R$ curves forwards (see Figure~\ref{fig:NS:GR:postTOV}) or backwards (see Figure~\ref{fig:NS:GR:postTOV2}). Therefore, the degeneracy between modifying gravity and modifying EOS is evident.

\section{MODIFIED GRAVITY: NON-PERTURBATIVE REGIME}

While the post-TOV formalism covers a variety of modified gravity theories, it fails to describe the non-perturbative behaviors that could be triggered by the inner strong gravitational field of NSs in some alternative gravity theories \cite{Damour:1993hw, Damour:1996ke, Glampedakis:2015sua, Shao:2017gwu}. 
``Spontaneous scalarization'' in the Damour \& Esposito-Far\`ese (DEF) gravity is an outstanding example \cite{Damour:1993hw, Damour:1996ke, Shao:2017gwu}.\footnote{In binary NS systems, a related phenomenon called ``dynamical scalarization'' is closely relevant to 
binary NS mergers
in the new field of gravitational-wave astrophysics \cite{Barausse:2012da, Shibata:2013pra, Sennett:2016rwa, Sennett:2017lcx}.} We will use NSs in the DEF gravity as an example for this section; for more examples, we refer the readers to the review by \citet{Doneva:2017jop} and references therein.

\subsection{An example: NSs in the DEF scalar-tensor gravity}

In the DEF gravity, a new scalar field, $\varphi$, is introduced with non-minimal couplings \cite{Damour:1992we, Damour:1993hw}. For the current content, it is easier to discuss in the Einstein frame, where the action for the geometry takes the Hilbert-Einstein form,
\begin{equation}
  \label{eq:action:STG}
  S = \frac{1}{16\pi G_*} \int \rmd^4 x \sqrt{-g_*} \left[ R_* - 2 g_*^{\mu\nu} \partial_\mu \varphi \partial_\nu \varphi - V\left( \varphi \right) \right] + S_{\rm matter} \left[ \psi; A^2\left( \varphi \right) g^*_{\mu\nu} \right] \,,
\end{equation}
where ``$*$'' means that we are in the Einstein frame; $G_*$ is the bare gravitational constant, and $V(\varphi)$ is the potential for the scalar field and we take it to be zero for simplicity. 
The most notable point in the action (\ref{eq:action:STG}) is the non-universal coupling of matter fields to the geometry (namely the metric $g_{\mu\nu}^*$) in $S_{\rm matter} \left[ \psi; A^2\left( \varphi \right) g^*_{\mu\nu} \right]$. Here the conformal factor $A\left( \varphi \right)$ is a function of $\varphi$ which can be spacetime-dependent. Consequently, for objects that source the scalar field, equivalence principle breaks down, and these objects do not follow the geodesics of $g_{\mu\nu}^*$. It is a manifestation of the strong equivalence principle violation, and has a deeper impact to the nature of gravitation \cite{Will:2014kxa, Berti:2015itd, Shao:2016ezh}. In addition, the divergence of $T^*_{\mu\nu}$ does not vanish.

For a spherically symmetric and stationary spacetime produced by a NS, the following metric was used \cite{Damour:1993hw} (here, and only in this place, $\varphi$ is a coordinate for the azimuthal angle, not to be confused with the scalar field),
\begin{equation}
  \label{eq:metric:STG}
  \rmd s_*^2 = - e^{\nu(r)} \rmd t^2 + \frac{\rmd r^2}{1 - 2G_* m(r)/r} + r^2 \left( \rmd \theta^2 + \sin^2 \theta \rmd \varphi^2 \right) \,.
\end{equation}

With the help of the field equations, a set of first-order differential equations were obtained by \citet{Damour:1993hw} for the structure of NSs,
\begin{eqnarray}
  \frac{\rmd \varphi}{\rmd r} &=& \psi \,, \\
  \frac{\rmd \psi}{\rmd r} &=&  \frac{4\pi G_*A^4(\varphi)}{1-2G_* m/r} \left[ \alpha(\varphi) \left( \epsilon - 3 p\right) + r \psi\left( \epsilon - p \right)  \right] - 2\frac{ 1- G_*m/r}{  1-2G_* m/r } \frac{\psi}{r} \,,\\
  \frac{\rmd m}{\rmd r} &=& 4\pi r^2 A^4(\varphi) \epsilon + \frac{r^2 \psi^2}{2G_*}  (1 - 2G_*m/r)  \,,  \\
  \frac{\rmd \nu}{\rmd r} &=&  \frac{8\pi G_* r A^4(\varphi) p}{1-2G_*m/r} + r\psi^2 + \frac{2 G_* m}{r^2\left( 1-2G_* m/r \right)} \,, \\
  \frac{\rmd p}{\rmd r} &=& - \left( \epsilon+ p \right) \left[   \frac{4\pi G_*r A^4(\varphi) p}{1-2G_*m/r} + \frac{1}{2}r \psi^2 + \frac{G_*m}{r^2(1-2G_*m/r)} + \alpha(\varphi) \psi \right] \,,
\end{eqnarray}
where $\alpha(\varphi)$ is defined by,
\begin{equation}
  \label{eq:alpha}
  \alpha(\varphi) \equiv \frac{\partial \ln A(\varphi)}{\partial \varphi} \,.
\end{equation}
There is a subtle point in above equations (in contrast to those in GR). The energy density and pressure ($\epsilon$ and $p$) are Jordan-frame/physical-frame variables, which were denoted as $\tilde \epsilon$ and $\tilde p$ in Refs.~\cite{Damour:1993hw, Damour:1996ke}. It means that their values measured in laboratories should be used.

In the following we will restrict ourselves to the DEF parameterization $\alpha(\varphi) = \beta_0 \varphi$, or equivalently,
\begin{equation}
  A(\varphi) = \exp \left( \frac{1}{2} \beta_0 \varphi^2 \right) \,.
\end{equation}
This is the simplest parameterization that reproduces significant strong-field deviations from GR.
We assume that the asymptotic value for $\varphi$ at spatial infinity is $\varphi_0$, and we denote $\alpha_0 \equiv \alpha(\varphi_0) = \beta_0 \varphi_0$ \cite{Damour:1993hw}. Therefore, the DEF scalar-tensor gravity is only described by two extra parameters, $\alpha_0$ and $\beta_0$ (or equivalently, $\varphi_0$ and $\beta_0$).
As we will see, $\alpha_0$ only smooths the non-perturbative transition behaviors, while
$\beta_0$ is the real game player \cite{Damour:1996ke, Sennett:2017lcx} that controls the critical point where the ``phase transition'' of spontaneous scalarization happens \cite{Damour:1993hw}.

The integration of the modified TOV equations in the DEF theory is similar to that in GR. Given a central energy density $\epsilon^{(c)}$ and the scalar field value at the center of a NS $\varphi^{(c)}$, one easily solves the above first-order differential equations \cite{Damour:1996ke}. In the case that one wants to fix the asymptotic value for the scalar field $\varphi_0$, a shooting algorithm and some iterations are needed \cite{Damour:1996ke}. 

\begin{figure}[ht]
  \centerline{\includegraphics[width=9cm]{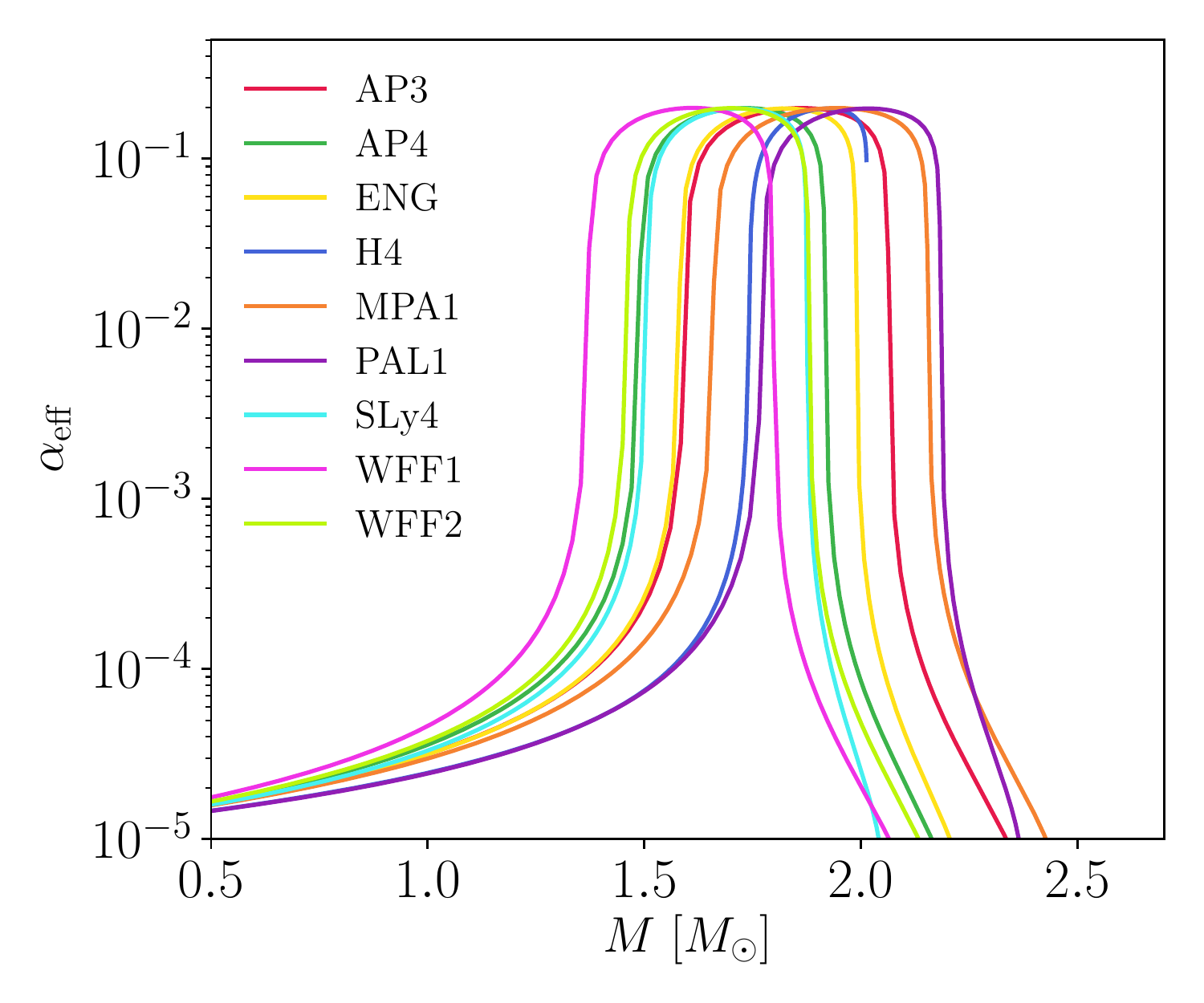}}
  \caption{\label{fig:scalar:charge} The effective scalar coupling in the DEF theory with $\left| \alpha_0 \right| = 10^{-5}$ and $\beta_0=-4.5$.}
\end{figure}

In addition to the metric and the matter distribution, the configuration for the scalar field is also obtained from solving the modified TOV equations. The effective scalar coupling for the NS is defined as,
\begin{equation}
  \label{eq:eff:scalar}
  \alpha_{\rm eff} \equiv \frac{\partial \ln M}{\partial \varphi_0} \,.
\end{equation}
\citet{Damour:1993hw} discovered that in the DEF theory, when $\beta_0 \lesssim
-4.5$, a non-perturbative phenomenon happens. After reaching a critical value of NS's compactness, a sudden increase by orders of magnitude in the effective scalar coupling $\alpha_{\rm eff}$ is observed. 
Such an increase introduces a large gravitational-wave dipole radiation (in addition to the
canonical quadrupole radiation in GR) in an asymmetric binary system, thus it
can be well constrained by the observations of binary pulsars
\cite{Wex:2014nva, Anderson:2019eay} or binary NS inspirals \cite{Will:1994fb,
Shao:2017gwu, Abbott:2018lct}. In Figure~\ref{fig:scalar:charge} we show the
effective scalar coupling of NSs as a function of their gravitational mass when
$\left| \alpha_0 \right| = 10^{-5}$ and $\beta_0=-4.5$. The spontaneous
scalarization is obvious for these curves. Nevertheless, it is interesting to
observe that, the non-perturbative phenomenon happens at different masses for different EOSs \cite{Shibata:2013pra, Shao:2017gwu}. 
For example, the EOS {\sf WFF1} has spontaneous scalarization at a relatively low mass, while the EOS {\sf PAL1} has spontaneous scalarization at a quite high mass. Different binary pulsar systems have different strength to probe this phenomenon for different EOSs \cite{Shao:2017gwu}.
For NSs with a very large mass, the EOS becomes ultra-relativistic, and the effective scalar coupling decreases (see \citet{Damour:1996ke} and \citet{EspositoFarese:2004cc} for more details).

\begin{figure}[ht]
  \centerline{\includegraphics[width=9cm]{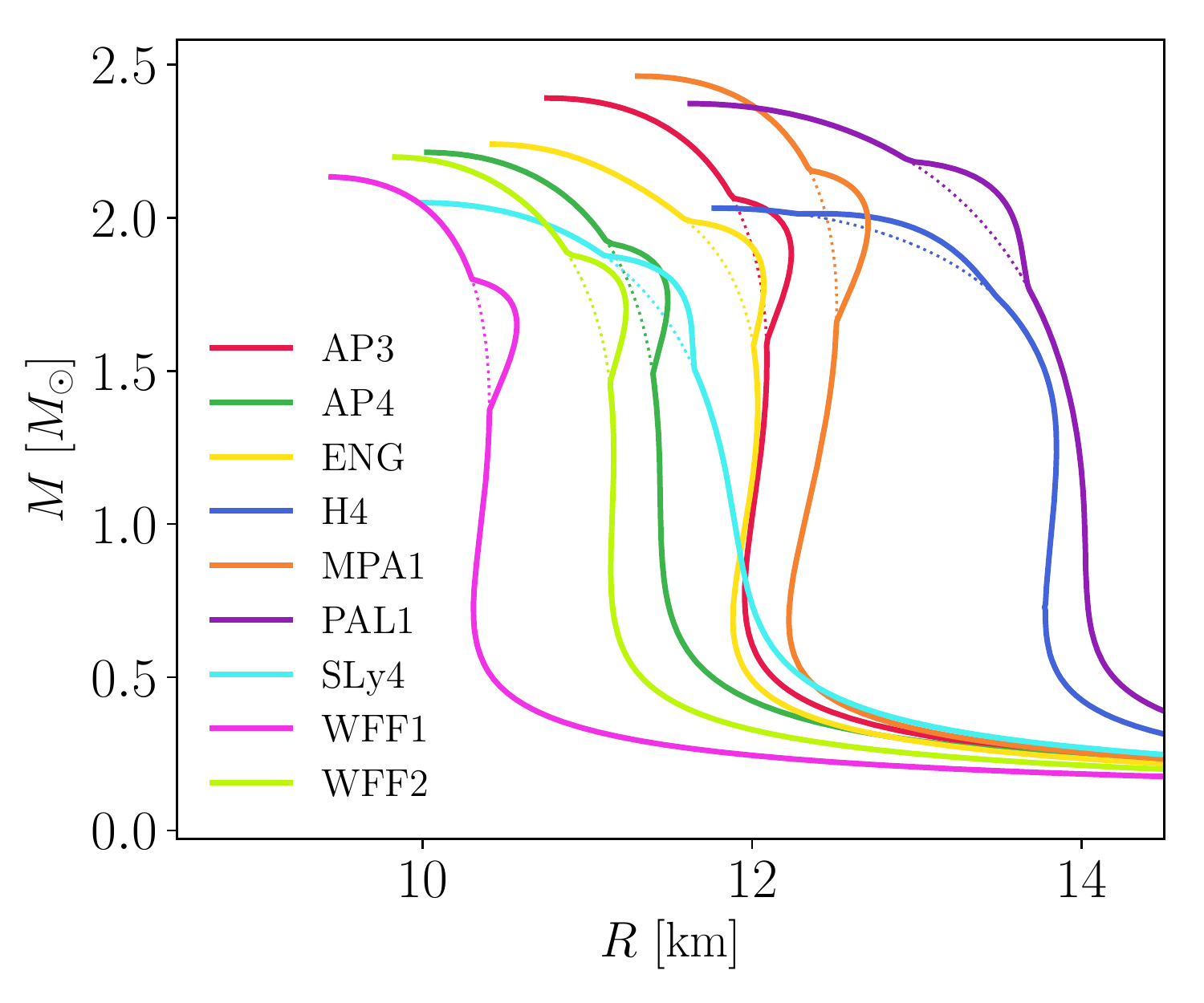}}
  \caption{\label{fig:NS:MR:STG} NSs' mass-radius relation in GR (dotted lines) and a DEF theory with $\left| \alpha_0 \right| = 10^{-5}$ and $\beta_0 = -4.5$ (solid lines).}
\end{figure}

We show the mass-radius relation for NSs in the DEF theory in
Figure~\ref{fig:NS:MR:STG} for $\left| \alpha_0 \right| = 10^{-5}$ and $\beta_0 = -4.5$. It is interesting to observe that (1) for most of
the parameter space, the mass-radius relation turns out to be very close to
that of GR; (2) there are ``bumps'' that appear for some masses for different
EOSs. These bumps are resulting from the non-perturbative behaviors, and they
are very hard, if ever possible, to be captured by the post-TOV equations
mentioned in the last section \cite{Glampedakis:2015sua}. The bumps make NSs
with a given mass become larger in radius, compared with that of GR. This is
a very distinct feature for spontaneously scalarized NSs. It can probably
play a role that any changes in the EOS could not mimic.

\section{DISCUSSIONS}

The properties of the high-density low-temperature nuclear matters, that
compose the NSs, carry very important knowledge for the community of nuclear
physics and astrophysics. It will tell us profound answers related to the
color-confinement in the quantum chromodynamics, the origin of mass, and the
evolution of our universe spanning from large (early cosmology) to small
(``frozen'' stars).

However, it is challenging to constrain the supranuclear EOS above several
times of the nuclear density from laboratory experiments
\cite{Lattimer:2015nhk}. {\it Extrapolation in science to some unknown regime
is not an easy task!} The input from astrophysics, especially from NS observations,
is valuable and complementary to what can be achieved on colliders. Most of valuable information comes from binary pulsar timing
observations \cite{Wex:2014nva}, X-ray observations for ``hot spots'' on the NS surface \cite{Watts:2014tja,
Watts:2018iom}, and, very recently, binary NS observations from gravitational waves
and the subsequent electromagnetic followups \cite{TheLIGOScientific:2017qsa,
Monitor:2017mdv}.

In this paper we discuss one caution when extracting the EOS information from
NSs, namely, the degeneracy with the not-so-well-tested gravity theory in the
strong gravitational field of NSs. As we learned from the past, scales matter
in physics. The validity of GR at different scales, no matter length scales
or field strength scales, should be tested empirically \cite{Psaltis:2008bb}. The validity of GR
has been tested to some extent in the strong field but not yet fully \cite{Yunes:2016jcc}, so we
shall be careful to interpret the observations. As we showed in this paper,
for example, the mass-radius relation of NSs could be different when the
gravity is not GR. On the theoretical hand, unlike the quantum chromodynamics
at low energy, GR elegantly makes definite predictions for the gravity
behavior even in the not-so-well-tested strong-field regime. This gives us a
lot reasonable confidence, alleviating some of our concerns. But still, in
the spirit of physical science, empirical verification is needed eventually, because we still have alternative gravity theories that agree with existing observations while making different predictions from GR for NSs.

As a new era for next-generation astronomical facilities is coming close, we
have a great hope to investigate both the NSs' EOS and the strong-field
gravity with new telescopes and observatories. For example, (1) the upcoming
Square Kilometre Array (SKA) \cite{Kramer:2004hd, Shao:2014wja, Bull:2018lat}
will provide us much better timing sensitivity than ever before, and it will
allow a decent measurement (or even several measurements) of the moment of
inertia for NSs at a good precision \cite{Kehl:2016mgp}. It encodes important
information about the supranuclear EOS. (2) The enhanced X-ray Timing and
Polarimetry mission (eXTP) \cite{Zhang:2018edu,Watts:2018iom,
intZand:2018yct, DeRosa:2018aka}, led by the Chinese and European teams, will
allow a precise measurement of NS radii via modeling the X-ray flux from the
hot spots formed through the accretion. It will shrink the uncertain region
in the mass-radius plots. (3) Last but not least, though the current
searching for the remnant of GW170817 \cite{Abbott:2017dke, Abbott:2018hgk}
is still prevented from positive detections by the large detector noises,
future gravitational-wave detectors will take us to identify the merger
remnant and to identify the characteristic oscillation modes implicating the
EOS of supranuclear matters. In all the above mentioned modeling, the
possibility of a deviation from GR can, in principle, be included, to reflect
our empirical uncertainty in the strong-field gravity. Such a deviation can
be fit together with the EOS. Being said, there is still a lot theoretical work remaining to be done.

In summary, most of the current approaches to study the supranuclear matters
use an implicit assumption that GR describes the strong gravitational field
inside NSs. This is not empirically verified to a safe precision. Therefore,
we shall at least keep a caution and work with this uncertainty. The
knowledge of EOSs is to be earned in the hard way, and a deeper understanding will happen in the near future.

\section{ACKNOWLEDGMENTS}
We thank Nils Andersson, Zhoujian Cao, James Lattimer, Bao-An Li, Renxin Xu, and Bing Zhang for helpful discussions during the Xiamen-CUSTIPEN Workshop on the EOS of Dense Neutron-Rich Matter in the Era of Gravitational Wave Astronomy. We are grateful to James Lattimer for providing us with tabulated data for neutron-star
equations of state. This work was supported by the National Science
Foundation of China (11721303), and XDB23010200.
\nocite{*}


%

\end{document}